\documentclass[twocolumn,prx,twocolumn,superscriptaddress]{revtex4-2}
 \usepackage[utf8]{inputenc}
\usepackage{color}
\usepackage{tikz}

\usepackage{lipsum}

\usepackage{upgreek}
\usepackage[normalem]{ulem}
\usepackage{mathtools}
\usepackage{datetime}
\usepackage{cancel}

\usepackage[all,cmtip,knot]{xy}
\usepackage{xypic}
\usepackage{braids}

\usepackage{amsmath}
\usepackage{amssymb}
\usepackage{graphicx}
\usepackage{tikz}
\usepackage{amsmath,bm,amssymb,latexsym,galois,amsthm,euscript,dsfont}
\usepackage{amsmath,bm,amssymb,latexsym,galois,amsthm,euscript}
\usepackage[all,cmtip,knot]{xy} 
\usepackage{mathrsfs}
\usepackage[unicode=true,bookmarks=true,bookmarksnumbered=false,bookmarksopen=false,breaklinks=false,pdfborder={0 0 1},backref=false,colorlinks=true]{hyperref}
\usepackage{algorithm}
\usepackage{algorithmic}

\hypersetup{
    colorlinks,
    linkcolor={red!60!black},
    citecolor={blue!90!black},
    urlcolor={blue!90!black}
}

\makeatletter

\usepackage{amsfonts}\usepackage{tabularx}\usepackage{dcolumn}\usepackage{bm}\usepackage{graphicx}\usepackage{epstopdf}

\newcommand{\orcidd}[1]{\href{https://orcid.org/#1}{\includegraphics[width=8pt]{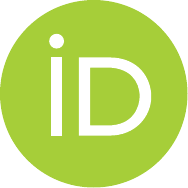}}}

\theoremstyle{remark}
\newtheorem{theorem}{Theorem}

\newtheorem{definition}{Definition}

\newtheorem{example}{Example}

\makeatletter
\newsavebox{\@brx}
\newcommand{\llangle}[1][]{\savebox{\@brx}{\(\m@th{#1\langle}\)}%
  \mathopen{\copy\@brx\kern-0.5\wd\@brx\usebox{\@brx}}}
\newcommand{\rrangle}[1][]{\savebox{\@brx}{\(\m@th{#1\rangle}\)}%
  \mathclose{\copy\@brx\kern-0.5\wd\@brx\usebox{\@brx}}}
\makeatother

\newcommand\Tr   {\operatorname{Tr}}

\usepackage{bm,latexsym,galois,euscript,dsfont}

\newcommand{\Acal}{\mathcal{A}}

\newcommand{\Bcal}{\mathcal{B}}

\newcommand{\State}{\mathsf{State}}
\newcommand{\Effect}{\mathsf{Effect}}
\newcommand{\Measurement}{\mathsf{Measurement}}

\newcommand{\Herm}{\mathsf{Herm}}
\newcommand{\Jtilde}{\tilde{J}}

\hypersetup{
    colorlinks,
    linkcolor={gray!80!black},
    citecolor={blue!90!black},
    urlcolor={gray!60!black}
}

\makeatletter

\usepackage{tikz}
\usetikzlibrary{positioning,intersections}
\usetikzlibrary{calc}
\usetikzlibrary{shapes.geometric}
\usepackage{braids}
\usepackage{tikz-cd}

\makeatother

\begin{document}

\title{The spatiotemporal doubled density operator: a unified framework for analyzing spatial and temporal quantum processes}

\author{Zhian Jia\orcidd{0000-0001-8588-173X}}
\email{giannjia@foxmail.com}
\affiliation{Centre for Quantum Technologies, National University of Singapore, Singapore 117543, Singapore}
\affiliation{Department of Physics, National University of Singapore, Singapore 117543, Singapore}

\author{Dagomir Kaszlikowski}
\email{phykd@nus.edu.sg}
\affiliation{Centre for Quantum Technologies, National University of Singapore, Singapore 117543, Singapore}
\affiliation{Department of Physics, National University of Singapore, Singapore 117543, Singapore}

\begin{abstract}
The measurement statistics for spatial and temporal quantum processes are produced through distinct mechanisms. Measurements that are space-like separated exhibit non-signaling behavior. However, time-like separated measurements can only result in one-way non-signaling, as the past is independent of the future, but the opposite is not true.
This work presents the doubled density operator as a comprehensive framework for studying quantum processes in space-time. It effectively captures all the physical information of the process, with the measurement and Born rule showing uniformity for both spatial and temporal cases.
We demonstrate that the equal-time density operator can be derived by performing a partial trace operation on the doubled density operator. Furthermore, the temporality of the quantum process can be detected by conducting a partial trace operation on either the left or right half of the doubled density operator.

\end{abstract}
\maketitle

\emph{Introduction.} --- 
Spatiotemporal quantum correlations exhibit significant potential for applications in various fields, including quantum chaos \cite{stockmann2000quantum, hosur2016chaos, Bertini2019exact, foligno2023temporal}, quantum stochastic processes \cite{Milz2021quantum}, quantum games \cite{gutoski2007toward, bell1964,Brunner2014RMP, leggett1985quantum, emary2013leggett}, quantum retrodiction \cite{Leifer2013toward,parzygnat2022axioms}, and other related domains. However, the treatment of space and time in standard quantum mechanics differs significantly from that in classical mechanics, where the two are usually considered on equal footing and unified under the concept of a space-time manifold.
While the probability distribution of particles in space can be obtained from the wavefunction using the Born rule \cite{born1926wellenmechanik}, no such notion exists for probability distribution over time.
In the Schrödinger (or more general, Dirac) equations, time is considered as a parameter that governs the evolution of the wavefunction \cite{Schrodinger1925an,dirac1928quantum}.
The goal of the modern quantum theory of gravity is to achieve a unified treatment of space and time, which necessitates the introduction of a quantized space-time manifold \cite{rovelli2004quantum}.
An alternative perspective involves examining quantum information processes that are implemented on a designated space-time lattice. In such a scenario, a temporal quantum process can be viewed as the preparation of a quantum system and the implementation of a sequence of measurements that yield a probability distribution $p(a_{t_n}, a_{t_{n-1}},\cdots,a_{t_0}|M_{t_n}, M_{t_{n-1}},\cdots,M_{t_0})$, wherein random variables are labeled by time coordinates \cite{leggett1985quantum,emary2013leggett}. Conversely, a spatial quantum process consists of preparing a multipartite state and measuring the local degrees of freedom to obtain a probability distribution $p(a_{x_m}, a_{x_{m-1}},\cdots,a_{x_1}|M_{x_m}, M_{x_{m-1}},\cdots,M_{x_1})$, with random variables labeled by space coordinates \cite{bell1964,Horodecki2009quantum,Brunner2014bell}.
From a quantum information perspective, it is possible to integrate both spatial multipartite quantum processes and temporal multi-time quantum processes into a unified framework and consider more general space-time quantum information processes.
This approach offers an alternative method for treating space and time equally and has gained considerable attention in recent years \cite{oreshkov2012quantum,griffiths1984consistent,cotler2016entangled,gutoski2007toward,cotler2018superdensity,Aharonov2009multi,fitzsimons2015quantum,jia2023quantumspace,fullwood2023quantum,fullwood2022quantum,liu2023quantum,hardy2005probability,Milz2021quantum}.
Various proposals for space-time states have been put forth, including process matrix \cite{oreshkov2012quantum}, consistent history \cite{griffiths1984consistent}, entangled history \cite{cotler2016entangled}, quantum-classical game \cite{gutoski2007toward}, superdensity operator \cite{cotler2018superdensity}, multi-time state \cite{Aharonov2009multi}, pseudo-density operator \cite{fitzsimons2015quantum}, and so on. Different approaches have their own advantages.

This work aims to address the fundamental question of whether a unified framework can be established that incorporates both spatial and temporal quantum processes, such that: (i) Their state, measurement, and Born rule are expressed in a uniform form; (ii) The equal time density operators can be recovered as the reduced states; (iii) The existence of temporality can be detected directly for the space-time state.
We propose the \emph{doubled density operator} formalism as a potential candidate to address this question, where the local Hilbert space for a given space-time event is doubled, the measurement for spatial and temporal are of the same form, and both the spatial and temporal quantum behavior can be obtained via the same Born rule.
The doubled density operator also provides a framework to investigate space-time quantum correlations. 

\begin{figure}
    \centering
    \includegraphics[scale=0.5]{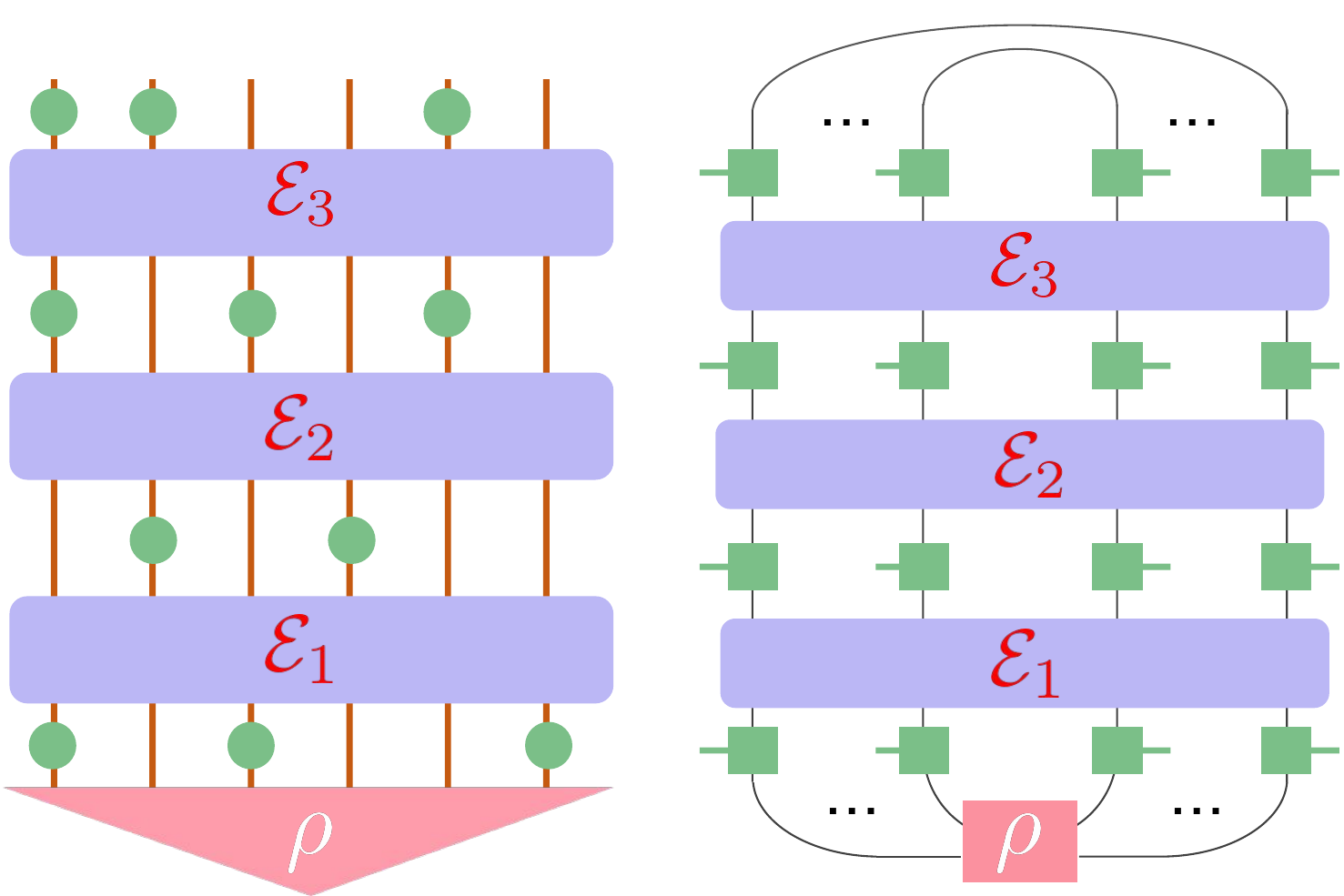}
    \caption{The presented figures provide a depiction of a general space-time quantum process and its corresponding correlation tensor. The left figure presents a quantum circuit representation of a space-time process, in which the green dot indicates the events that are of interest. On the other hand, the right figure demonstrates a tensor network representation of the correlation tensor associated with a general space-time doubled density operator. This tensor network depiction is useful in visualizing and analyzing the intricate structure of the correlation tensor, facilitating a deeper understanding of the underlying quantum process.}
    \label{fig:DDO}
\end{figure}

\emph{Doubled density operator.} ---
Suppose $N$ parties are distributed across space-time, with each able to implement arbitrary quantum measurements over their local quantum systems. 
The term 'party' is often used in quantum games to describe quantum correlations, but for the purposes of this discussion, we will also refer to them as 'events'.
For a given event we double the corresponding space as $\mathcal{H}_{i,L}\cong \mathcal{H}_{i,R}\cong \mathcal{H}_{i}$.
We assume that the local space has arbitrary $d$ dimensions and that the measurements are represented by generalized Pauli operators, also known as Hilbert-Schmidt basis operators $\{\sigma_{\mu}\}_{\mu=0}^{d-1}$. These operators are Hermitian and satisfy the following conditions: (i) $\sigma_0$ is the identity operator $\mathds{I}$; (ii) the trace of $\sigma_j$ is zero for all $j\geq 1$; and (iii) the matrices are orthogonal, meaning that $\Tr(\sigma_\mu \sigma_\nu)=d\delta_{\mu\nu}$. See Ref. \cite{wei2022antilinear} for an explicit matrix form we will use.
The generalized Pauli operators indeed form a basis for the space of $d\times d$ Hermitian operators $\mathsf{Herm}(\mathbb{C}^d)$, when regarded as a real vector space. They also form a basis for the space of all $d\times d$ linear operators $\mathcal{B}(\mathbb{C}^d)$, when regarded as a complex vector space. We will also use the Hilbert-Schmidt inner product $\langle X,Y\rangle = \Tr (X^{\dagger}Y)$ for operators.
Our objective is to construct a mathematical representation of states, measurements, and the Born rule within a given $C^*$-algebra $\mathcal{B}(\mathcal{H}_L\otimes \mathcal{H}_R)$, where $\mathcal{H}_L=\mathcal{H}_R=\otimes_{i=1}^N \mathcal{H}_i$.
To this end, we introduce the \emph{space-time doubled correlation tensor} (DCT) $T^{\mu_1,\cdots,\mu_N;\nu_1,\cdots,\nu_N}$ where $\mu_i,\nu_i=0,\cdots, d-1$ are left and right indices for $i$-th event.

\begin{definition}[spatiotemporal doubled correlation tensor]\label{def:STtensor}
A complex-valued tensor $T^{\mu_1,\cdots,\mu_N;\nu_1,\cdots,\nu_N}$ with $2N$ indices  is called a space-time correlation tensor if and only if it satisfies the following conditions
\begin{enumerate}
    \item $T$ is Hermitian with respect to left and right indices
    \begin{equation}
(T^{\nu_1,\cdots,\nu_N;\mu_1,\cdots,\mu_N})^*=T^{\mu_1,\cdots,\mu_N;\nu_1,\cdots,\nu_N}.
    \end{equation}
    \item $T$ is positive semidefinite with respect to left and right indices, viz., for any complex tensor $X_{\alpha_1,\cdots,\alpha_N}$
    \begin{equation}
        \sum_{\mu_i;\nu_i} X_{\mu_1,\cdots,\mu_N}^* T^{\mu_1,\cdots,\mu_N;\nu_1,\cdots,\nu_N}X_{\nu_1,\cdots,\nu_N} \geq 0,
    \end{equation}
    and the equality holds only when $X_{\alpha_1,\cdots,\alpha_N}\equiv 0$.
     \item $T^{0,\cdots,0;0,\cdots,0}=1$.
\end{enumerate}
We denote the set of all $N$-event doubled correlation tensors as $\mathsf{DCT}(N)$. It's clear that $\mathsf{DCT}(N)$ is closed under the convex combination.
\end{definition}

The spatial doubled correlation tensor for a $N$-particle state is defined as 
\begin{equation}\label{eq:stensor}
    T^{\mu_1,\cdots,\mu_N;\nu_1,\cdots,\nu_N}=\Tr[ (\otimes_{i=1}^N \sigma_{\mu_i}) \rho (\otimes_{i=1}^N \sigma_{\nu_i})]. 
\end{equation}
The $n$-step temporal doubled correlation tensor for a single qudit state $\rho$ is defined as 
\begin{align}
     &T^{\mu_1,\cdots,\mu_N;\nu_1,\cdots,\nu_N} \nonumber\\
     =&\Tr [\sigma_{\mu_N} \mathcal{E}_{N-1}(\cdots \mathcal{E}_1(\sigma_{\mu_1}\rho\sigma_{\nu_1}) \cdots)\sigma_{\nu_N}], \label{eq:ttensor}
\end{align}
where $\mathcal{E}_i$ are quantum channels that are completely positive trace-preserving (CPTP) maps.
A general space-time doubled correlation tensor is of the form

\begin{widetext}
\begin{align}
T^{\mu^0_{1},\cdots,\mu^0_{m_0},\cdots,\mu^n_{1},\cdots,\mu^n_{m_n};\nu^0_{1},\cdots,\nu^0_{m_0},\cdots,\nu^n_{1},\cdots,\nu^n_{m_n}}=  
\Tr \left[ (\otimes_{j_n} \sigma_{\mu^n_{j_n}})  \mathcal{E}_n (\cdots  \mathcal{E}_1((\otimes_{j_0} \sigma_{\mu^0_{j_0}})\rho  (\otimes_{l_0} \sigma_{\nu^0_{l_0}})) \cdots)  (\otimes_{l_n} \sigma_{\nu^n_{l_n}}) \right], \label{eq:sttensor}
\end{align}
\end{widetext}
where at each time step we choose several local degrees of freedom to measure, see Fig.~\ref{fig:DDO} for an illustration.
If at each time step of a space-time quantum process, we measure all possible local spatial degrees of freedom, the resulting space-time correlation tensor is called \emph{information-complete}.
Notice that Eqs.~\eqref{eq:stensor} and \eqref{eq:ttensor} are special forms of Eq.~\eqref{eq:sttensor}.
All the above tensors satisfy the axioms in definition \ref{def:STtensor}, see supplemental material for proofs.

Since the density operator is defined in $\mathcal{B}(\mathcal{H}_L)$ (or equivalently $\mathcal{B}(\mathcal{H}_R)$), in this sense, we extend the Hilbert space by a factor of two and define the space-time state in $\mathcal{B}(\mathcal{H}_L\otimes \mathcal{H}_R)$, and refer to the resulting states as the \emph{doubled density operator} (DDO).

\begin{definition}
For a given  spatiotemporal DCT $T^{\{\mu_i\};\{\nu_j\}}$, the corresponding DDO is defined as follows
\begin{align}
    W=&\frac{1}{d^{2N}}\sum_{\mu_i;\nu_i} T^{\{\mu_i\};\{\nu_j\}} (\otimes_{i=1}^N \sigma_{\mu_i}) \otimes (\otimes_{j=1}^N \sigma_{\nu_j}).
\end{align}
It's clear that $W$ is in general not Hermitian and has unit trace $\Tr W=1$. For a given event set $\Acal$, we denote the set of all DDOs of $\Acal$ as $\mathsf{DDO}(\Acal)$.
\end{definition}

Another natural condition to impose on the DDO is that its one-event reduced DDO always corresponds to a density operator. This condition ensures that the one-event reduced DCT takes the form that will be described in Eq.~\eqref{eq:singleT} later.

Notice that via the vector map $|\bullet\rrangle: \mathcal{B}(\mathcal{H})\to \mathcal{H}\otimes \mathcal{H}$ defined by $||i\rangle \langle j|\rrangle=|i\rangle \otimes |j\rangle$ (the dual map $\llangle \bullet |$ can be defined similarly), the space-time doubled density operator $W$ can be mapped to a superdensity density operator $\varrho_W$, which is introduced in Ref. \cite{cotler2018superdensity}.
The left and right components are mapped as follows: $\sigma_{\mu_i}\mapsto |\sigma_{\mu_i}\rrangle$ and $\sigma_{\nu_i}\mapsto \llangle \sigma_{\nu_i}|$. 
However, since $\rho_W$ is a positive semidefinite operator with unit trace for any space-time DCT defined in definition \ref{def:STtensor}, the temporal, or causal, aspects of the system cannot be observed directly, unlike the pseudo-density operator for which negativity of state indicates the existence of temporal correlation \cite{fitzsimons2015quantum}.
Moreover, for the pseudo-density operator, the usual spatial density operator at a specific time can be obtained by simply taking the partial trace, but for the superdensity operator, the reduced state is still a superdensity operator.
The pseudo-density operator suffers from the problem that for spatial and temporal cases, there is no unified Born rule. 
As we shall observe subsequently, the DDO effectively integrates the benefits of both approaches and surmounts some of the challenges mentioned earlier. Therefore, it provides a suitable and comprehensive framework for investigating space-time quantum states and space-time quantum correlations.

To illustrate, let us start with the fundamental building block, the case of a single event.
Consider a single-qubit state, where the correlation tensor for $\rho=\sum_{\mu=0}^3 c_{\mu}\sigma_{\mu}/2$ is of the form    $T^{\mu;\nu}=\Tr (\sigma_{\mu}\rho \sigma_{\nu})$.
A direct calculation gives
\begin{equation} \label{eq:singleT}
    T^{\mu ;\nu}=\left(\begin{array}{cccc}
c_0 & c_1 & c_2 & c_3 \\
c_1 & c_0 & -i c_3 & i c_2 \\
c_2 & i c_3 & c_0 & -i c_1 \\
c_3 & -i c_2 & i c_1 & c_0
\end{array}\right),
\end{equation}
which is a $4\times 4$ Hermitian matrix.
The corresponding doubled density operator is of the form
\begin{equation}\label{eq:singleW}
    W=\frac{1}{2^2}\sum_{\mu,\nu=0}^3 T^{\mu;\nu} \sigma_{\mu}\otimes \sigma_{\nu}.
\end{equation}
Notably, the matrix $W$ is non-Hermitian, but we have $\Tr W=1$. Furthermore, we have $\rho=\Tr_{L} W=\Tr_{R} W$, hereinafter $L$ and $R$ refer to the left and right halves of the state.
It's convenient to use the tensor network to represent correlation tensor $T^{\mu;\nu}$ as
\begin{equation}
    T^{\mu;\nu}= \begin{aligned}
	\begin{tikzpicture}
			\draw[line width=.6pt,black] (-0.9,1.25) -- (-1.3,1.25);
               \draw[line width=.6pt,black] (0.9,1.25) -- (1.3,1.25);
                \draw[line width=0.6 pt, fill=gray, fill opacity=0.2] 
		(-0.3,0) -- (0.3,0) -- (0.3,0.5) -- (-.3,0.5) -- cycle;  
\draw[line width=.6pt,black]  plot [smooth,tension=0.6] 
coordinates {(-0.3,0.25) (-0.6,.5) (-.6,1) }; 
\draw[line width=.6pt,black]  plot [smooth,tension=0.6] 
coordinates {(0.3,0.25) (0.6,.5) (.6,1) }; 
\draw[line width=.6pt,black]  plot [smooth,tension=0.6] 
coordinates {(-0.5,1.5) (0,2) (.5,1.5) }; 
       \draw[line width=0.6 pt, fill=gray, fill opacity=0.2] 
		(-0.9,1) -- (-0.3,1) -- (-0.3,1.5) -- (-.9,1.5) -- cycle; 
   \draw[line width=0.6 pt, fill=gray, fill opacity=0.2] 
		(0.9,1) -- (0.3,1) -- (0.3,1.5) -- (.9,1.5) -- cycle; 
      \node[ line width=0.6pt, dashed, draw opacity=0.5] (a) at (0,0.25)
      {$\rho$};
    \node[ line width=0.6pt, dashed, draw opacity=0.5] (a) at (-0.6,1.25)
      {$\sigma_{\mu}$};
         \node[ line width=0.6pt, dashed, draw opacity=0.5] (a) at (0.6,1.25)
      {$\sigma_{\nu}$};
		\end{tikzpicture}
	\end{aligned}
\end{equation}
where each box corresponds to an operator, with the connections between boxes indicating matrix multiplication, and the dangling edges representing free indices $\mu$ and $\nu$ labeling the Pauli basis for operators.  See, e.g., Ref. \cite{Orus2014a}, for an introduction to the tensor network.

In the case of two events in space, the spatial doubled correlation tensor is 
\begin{equation}
    T^{\mu_1,\mu_2;\nu_1,\nu_2}=\Tr ((\sigma_{\mu_1}\otimes \sigma_{\mu_2}) \rho (\sigma_{\nu_1}\otimes \sigma_{\nu_2})),
\end{equation}
where $\rho$ is now a two-qubit state.
The doubled density operator is of the form
\begin{equation}\label{eq:twoqubitDDD}
 W=  \frac{1}{2^4} \sum_{\mu_1,\mu_2;\nu_1,\nu_2=0}^3   T^{\mu_1,\mu_2,\nu_1,\nu_2} (\sigma_{\mu_1}\otimes \sigma_{\mu_2}) \otimes (\sigma_{\nu_1}\otimes \sigma_{\nu_2}).
\end{equation}
The tensor network for correlation tensor is as follows
\begin{equation}
    T^{\mu_1,\mu_2;\nu_1,\nu_2}= \begin{aligned}
	\begin{tikzpicture}
			\draw[line width=.6pt,black] (-0.9,1.25) -- (-1.3,1.25);
               \draw[line width=.6pt,black] (0.9,1.25) -- (1.3,1.25);
              \draw[line width=.6pt,black] (-1.9,1.95) -- (-2.3,1.95);
               \draw[line width=.6pt,black] (1.9,1.95) -- (2.3,1.95);
                \draw[line width=0.6 pt, fill=gray, fill opacity=0.2] 
		(-0.3,0) -- (0.3,0) -- (0.3,0.5) -- (-.3,0.5) -- cycle;  
\draw[line width=.6pt,black]  plot [smooth,tension=0.6] 
coordinates {(-0.3,0.4) (-0.6,.5) (-.6,1) }; 
\draw[line width=.6pt,black]  plot [smooth,tension=0.6] 
coordinates {(0.3,0.4) (0.6,.5) (.6,1) }; 
\draw[line width=.6pt,black]  plot [smooth,tension=0.6] 
coordinates {(-0.5,1.5) (0,2) (.5,1.5) }; 
\draw[line width=.6pt,black]  plot [smooth,tension=0.6] 
coordinates {(-0.3,0.2) (-1.6,.5) (-1.6,1.7) }; 
\draw[line width=.6pt,black]  plot [smooth,tension=0.6] 
coordinates {(0.3,0.2) (1.6,.5) (1.6,1.7) }; 
\draw[line width=.6pt,black]  plot [smooth,tension=0.6] 
coordinates {(-1.5,2.2) (0,2.8) (1.5,2.2) }; 
       \draw[line width=0.6 pt, fill=gray, fill opacity=0.2] 
		(-0.9,1) -- (-0.3,1) -- (-0.3,1.5) -- (-.9,1.5) -- cycle; 
   \draw[line width=0.6 pt, fill=gray, fill opacity=0.2] 
		(0.9,1) -- (0.3,1) -- (0.3,1.5) -- (.9,1.5) -- cycle; 
         \draw[line width=0.6 pt, fill=gray, fill opacity=0.2] 
		(-1.9,1.7) -- (-1.3,1.7) -- (-1.3,2.2) -- (-1.9,2.2) -- cycle; 
   \draw[line width=0.6 pt, fill=gray, fill opacity=0.2] 
		(1.9,1.7) -- (1.3,1.7) -- (1.3,2.2) -- (1.9,2.2) -- cycle; 
      \node[ line width=0.6pt, dashed, draw opacity=0.5] (a) at (0,0.25)
      {$\rho$};
    \node[ line width=0.6pt, dashed, draw opacity=0.5] (a) at (-0.6,1.25)
      {$\sigma_{\mu_2}$};
         \node[ line width=0.6pt, dashed, draw opacity=0.5] (a) at (0.6,1.25)
      {$\sigma_{\nu_2}$};
         \node[ line width=0.6pt, dashed, draw opacity=0.5] (a) at (-1.6,1.95)
      {$\sigma_{\mu_1}$};
         \node[ line width=0.6pt, dashed, draw opacity=0.5] (a) at (1.6,1.95)
      {$\sigma_{\nu_1}$};
		\end{tikzpicture}
	\end{aligned}.
\end{equation}
The generalization to the $n$-qubit case is straightforward.
It is clear from the above equation that fixing $\mu_1=\nu_1=0$ gives us the correlation tensor for the doubled reduced density operator of the second qubit, and similarly for fixing $\mu_2=\nu_2=0$.
This relationship holds for the general $n$-qubit correlation tensor $T^{\mu_1,\cdots,\mu_n,\nu_1,\cdots,\nu_n}$, and all possible reduced doubled density operator correlation tensors can be obtained by setting the indices not contained in it to zero.
An intriguing observation is that if we set the right indices $\nu_1=\nu_2=0$ in the correlation tensor of the two-event spatial case, we obtain the correlation tensor for the usual Bloch representation of the two-qubit density operator \cite{wei2022antilinear}. For left indices, a similar result holds.
This means that the density operator can be recovered from the doubled density operator by taking the right or left reduced state.

For two temporal events, we assume that the input state is $\rho$ and the quantum channel is $\mathcal{E}$. The correlation tensor is defined as 
\begin{equation}
    T^{\mu_1,\mu_2;\nu_1,\nu_2}=\Tr (\sigma_{\mu_2} \mathcal{E}(\sigma_{\mu_1}\rho \sigma_{\nu_1}) \sigma_{\nu_2}).
\end{equation}
The corresponding temporal doubled density operator is of the same form as in Eq.~\eqref{eq:twoqubitDDD}.
In tensor network representation, we have
\begin{equation}
    T^{\mu_1,\mu_2;\nu_1,\nu_2}= \begin{aligned}
	\begin{tikzpicture}
			\draw[line width=.6pt,black] (-0.9,1.25) -- (-1.3,1.25);
               \draw[line width=.6pt,black] (0.9,1.25) -- (1.3,1.25);
               	\draw[line width=.6pt,black] (-0.9,2.65) -- (-1.3,2.65);
               \draw[line width=.6pt,black] (0.9,2.65) -- (1.3,2.65);
                \draw[line width=0.6 pt, fill=gray, fill opacity=0.2] 
		(-0.3,0) -- (0.3,0) -- (0.3,0.5) -- (-.3,0.5) -- cycle;  
\draw[line width=.6pt,black]  plot [smooth,tension=0.6] 
coordinates {(-0.3,0.25) (-0.6,.5) (-.6,1) }; 
\draw[line width=.6pt,black]  plot [smooth,tension=0.6] 
coordinates {(0.3,0.25) (0.6,.5) (.6,1) }; 
\draw[line width=.6pt,black] (-0.6,1.5) -- (-0.6,1.7);
\draw[line width=.6pt,black] (0.6,1.5) -- (0.6,1.7);
     \draw[line width=0.6 pt, fill=gray, fill opacity=0.2] 
		(-0.9,1.7) -- (0.9,1.7) -- (0.9,2.2) -- (-.9,2.2) -- cycle;
\draw[line width=.6pt,black] (-0.6,2.2) -- (-0.6,2.4);
\draw[line width=.6pt,black] (0.6,2.2) -- (0.6,2.4);
       \draw[line width=0.6 pt, fill=gray, fill opacity=0.2] 
		(-0.9,2.4) -- (-0.3,2.4) -- (-0.3,2.9) -- (-.9,2.9) -- cycle; 
   \draw[line width=0.6 pt, fill=gray, fill opacity=0.2] 
		(0.9,2.4) -- (0.3,2.4) -- (0.3,2.9) -- (.9,2.9) -- cycle; 
\draw[line width=.6pt,black]  plot [smooth,tension=0.6] 
coordinates {(-0.5,2.9) (0,3.3) (.5,2.9) }; 
       \draw[line width=0.6 pt, fill=gray, fill opacity=0.2] 
		(-0.9,1) -- (-0.3,1) -- (-0.3,1.5) -- (-.9,1.5) -- cycle; 
   \draw[line width=0.6 pt, fill=gray, fill opacity=0.2] 
		(0.9,1) -- (0.3,1) -- (0.3,1.5) -- (.9,1.5) -- cycle; 
      \node[ line width=0.6pt, dashed, draw opacity=0.5] (a) at (0,0.25)
      {$\rho$};
    \node[ line width=0.6pt, dashed, draw opacity=0.5] (a) at (-0.6,1.25)
      {$\sigma_{\mu_1}$};
         \node[ line width=0.6pt, dashed, draw opacity=0.5] (a) at (0.6,1.25)
      {$\sigma_{\nu_1}$};
          \node[ line width=0.6pt, dashed, draw opacity=0.5] (a) at (0,1.95)
      {$\mathcal{E}$};
          \node[ line width=0.6pt, dashed, draw opacity=0.5] (a) at (-0.6,2.65)
      {$\sigma_{\mu_2}$};
         \node[ line width=0.6pt, dashed, draw opacity=0.5] (a) at (0.6,2.65)
      {$\sigma_{\nu_2}$};
		\end{tikzpicture}
	\end{aligned}
\end{equation}
The generalization to the multi-time process is straightforward. Notice that in this case, the left and right reduced states are no longer density operators.
For instance, if $\mathcal{E}=\operatorname{id}$, $T^{0,0;\nu_1,\nu_2}=\Tr (\rho \sigma_{\nu_1}\sigma_{\nu_2})$ for which $T^{0,0;1,2}=i \Tr (\rho \sigma_z)$ is not real, thus cannot be a Bloch tensor of a density operator \cite{wei2022antilinear}.

Now consider the general space-time quantum process as shown in Fig.~\ref{fig:DDO}, which consists of $n+1$ time steps where $\mathcal{E}_1,\cdots,\mathcal{E}_n$ represent the channels between each time step. At each time step $t_k$, we choose to measure $m_k$ spatial degrees of freedom, and the corresponding left and right generalized Pauli measurements are denoted as $\sigma_{\mu^k_1},\cdots,\sigma_{\mu^k_{m_k}}$ and $\sigma_{\nu^k_1},\cdots,\sigma_{\nu^k_{m_k}}$ respectively. The corresponding space-time state for a total of $N=\sum_{k=0}^n m_k$ events is determined by the space-time doubled correlation tensor in Eq.~\eqref{eq:sttensor}.
The corresponding space-time doubled density operator is of the form
\begin{widetext}
  \begin{align}
    W=\frac{1}{d^{2N}}\sum_{\mu^i_{j_i};\nu^q_{l_q}} T^{\mu^0_{1},\cdots,\mu^0_{m_0},\cdots,\mu^n_{1},\cdots,\mu^n_{m_n};\nu^0_{1},\cdots,\nu^0_{m_0},\cdots,\nu^n_{1},\cdots,\nu^n_{m_n}}
     [\otimes_{k=0}^n (\otimes_{j_i=1}^{m
    _i} \sigma_{\mu^i_{j_i}})]\otimes [\otimes_{q=0}^n   (\otimes_{l_q=1}^{m
    _q} \sigma_{\nu^q_{l_q}})].
\end{align}  
\end{widetext}

If at each time step of a space-time quantum process, we measure all possible local spatial degrees of freedom, the resulting space-time state is called \emph{information-complete}. 
By taking the appropriate reduced state from the information-complete doubled density operator $W$, we can obtain all possible physical states that appear in a space-time quantum process. This includes the output state $\rho_{t_k}$ for each time step, as well as their reduced states.
    Tracing out all the parts of $W$ except for the $t_k$ step, we obtain a spatial doubled density operator $W(\rho_{t_k})$.
    Then, tracing out the right (or left) half of $W(\rho_{t_k})$, we obtain the usual density operator of $\rho_k$.
In this sense, the doubled density operator can be regarded as a space-time state.
By generalizing the Choi-Jamio{\l}kowski isomorphism  \cite{choi1975completely,jamiolkowski1972linear} to the generalized Pauli basis, we can obtain a closed-form expression for the general doubled density operator. 
It's useful to define the \emph{doubled generalized Jamio{\l}kowski matrix}  as (for single particle case) 
\begin{align}
     &\mathcal{J}^{\{\mu_i\};\{\nu_i\}}[\mathcal{E}_1,\cdots,\mathcal{E}_{N-1}]\nonumber\\
    =&\frac{1}{d}\sum_{\alpha=0}^{d-1}[\sigma_{\mu_N} \mathcal{E}_{N-1}(\cdots \mathcal{E}_1(\sigma_{\mu_1}\sigma_{\alpha}\sigma_{\nu_1}) \cdots)\sigma_{\nu_N}] \otimes \sigma_{\alpha}.
\end{align}
The doubled correlation tensor can be obtained by taking the trace of  $\mathcal{J}^{\{\mu_i\};\{\nu_i\}}[\mathcal{E}_1,\cdots,\mathcal{E}_{N-1}]$.
More interestingly, if we define 
\begin{equation}
    \mathcal{W}=\sum_{{\{\mu_i\};\{\nu_i\}}} \mathcal{J}^{\{\mu_i\};\{\nu_i\}}[\mathcal{E}_1,\cdots,\mathcal{E}_{N-1}] \otimes [(\otimes_{i}\sigma_{\mu_i})\otimes (\otimes_j \sigma_{\nu_j})]
\end{equation}
the doubled density operator is given by
\begin{equation}
W=    \Tr_{\mathcal{J}}   \mathcal{W} (\mathds{I}\otimes \rho \otimes \mathds{I} \otimes \cdots \otimes \mathds{I}),
\end{equation}
where the trace for $\mathcal{J}$ means that the trace is taking over the spaces that support $\mathcal{J}^{\{\mu_i\};\{\nu_i\}}[\mathcal{E}_1,\cdots,\mathcal{E}_{N-1}] $.
This gives us a closed-form expression for a general temporal DDO. The generalization to the space-time case is straightforward.
See supplemental material for a detailed discussion.

\emph{Measurement and spatiotemporal Born rule.} ---
We now introduce the concept of the \emph{doubled effect tensor} (DET) for \emph{doubled measurement} (DM), which can also be represented by a tensor with indices labeled by a generalized Pauli basis. By contracting the doubled correlation tensor and the effect tensor, we obtain the corresponding probability distribution for a given space-time quantum process.
This unifies the spatial and temporal quantum processes using the same definitions of states and measurement, and we have a unified generalized Born rule.

\begin{definition}[Space-time measurement]\label{def:measurement}
   For an $N$-event set $\Acal=\{1,\cdots,N\}$, we define the doubled measurement as follows:
   \begin{enumerate}
       \item The space-time local measurement for outputting $a_1,\cdots,a_N$ is of the form 
     $M_{a_1,\cdots,a_N}
     =(K^1_{a_1}\otimes\cdots \otimes K^N_{a_N})\otimes ({K^1_{a_1}}^{\dagger}\otimes\cdots \otimes {K^N_{a_N}}^{\dagger})$,
   where the operator $K_{a_i}^i$'s are required to make $F_{a_i}=(K^{i}_{a_i})^{\dagger}K^{i}_{a_i}$  local POVM elements, viz., $\sum_{a_i}(K^{i}_{a_i})^{\dagger}K^{i}_{a_i}=\mathds{I}$.
     \item The space-time joint measurement for outputting $a$ is of the form $M_a=K_a\otimes K_a^{\dagger}$ with $F_a=K_a^{\dagger}K_a$ the joint POVM, viz., $K_{a}$ jointly acts on $\mathcal{H}_1\otimes \cdots \otimes \mathcal{H}_N$.
   \end{enumerate}
   Notice that the local and joint measurements for the spatial and temporal cases are both of the same forms.
\end{definition}

\begin{figure}
    \centering
    \includegraphics[scale=0.35]{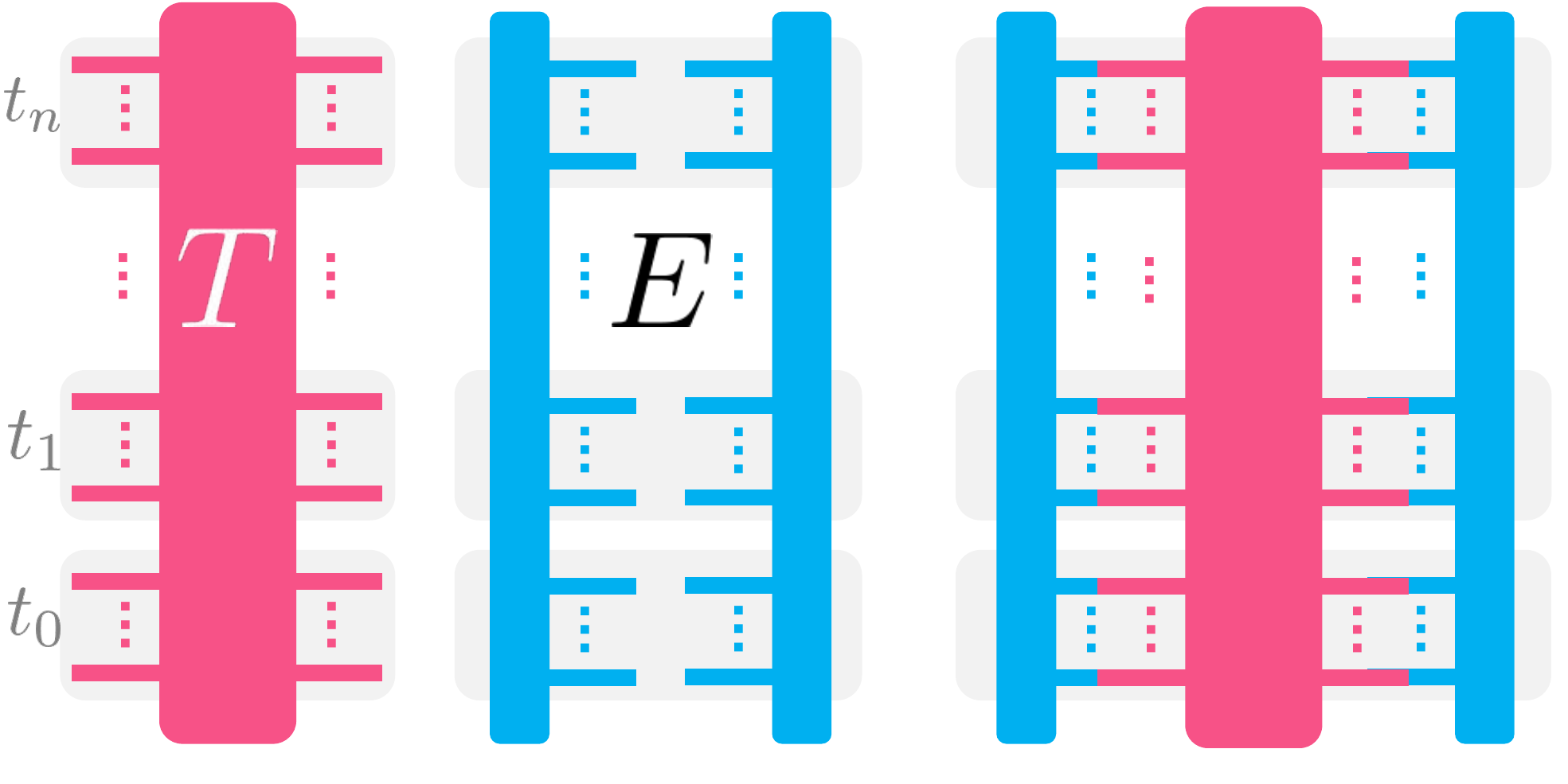}
    \caption{The depiction of the doubled correlation tensor for space-time state, doubled measurement tensor of quantum effect, and Born rule of the doubled density operator in the tensor network representation. Notice that the overall factor about dimensions of spaces has been omitted.}
    \label{fig:Born}
\end{figure}

Consider a single-event DDO given by Eq.~\eqref{eq:singleW}, the Born rule in quantum mechanics is of the form
\begin{equation}
    p(a)=\Tr(K_a \rho K_a^{\dagger})=\Tr (F_a \rho).
\end{equation}
The DM is of the form 
\begin{equation}
    M_a= K_a\otimes K_a^{\dagger}.
\end{equation}
The generalized Born rule is given by
\begin{equation}
    p(a)=\Pr (M_a, W)=\Tr (M_aW),
\end{equation}
which is easily checked to match well with the result of standard quantum mechanics.
It is also useful to introduce the tensor network representation of 
 DET $E_a^{\mu;\nu}=\Tr (M_a (\sigma_{\mu}\otimes \sigma_{\nu}))/2^2$ as follows,
\begin{equation}
   E^{\mu;\nu} =\frac{1}{2^2}\begin{aligned}
	\begin{tikzpicture}
			\draw[line width=.6pt,black] (0.6,0.25) -- (1,0.25);
               \draw[line width=.6pt,black] (1.9,.25) -- (1.5,.25);
               \draw[line width=.6pt,black] (0.3,-0.3) -- (.3,0);
               \draw[line width=.6pt,black] (2.2,-.3) -- (2.2,0);
\draw[line width=.6pt,black]  plot [smooth,tension=0.6] 
coordinates {(0.3,.5) (-0.2,0.7) (-.4,-0.2) (-.2,-1) (0.3,-.8)}; 
\draw[line width=.6pt,black]  plot [smooth,tension=0.6] 
coordinates {(2.2,.5) (2.7,0.7) (2.9,-0.2) (2.7,-1) (2.2,-.8)}; 
       \draw[line width=0.6 pt, fill=gray, fill opacity=0.2] 
		(0,0) -- (0.6,0) -- (.6,.5) -- (0,.5) -- cycle; 
   \draw[line width=0.6 pt, fill=gray, fill opacity=0.2] 
		(1.9,0) -- (2.5,0) -- (2.5,.5) --(1.9,0.5) -- cycle; 
         \draw[line width=0.6 pt, fill=gray, fill opacity=0.2] 
		(0,-0.8) -- (0.6,-0.8) -- (.6,-.3) -- (0,-.3) -- cycle; 
   \draw[line width=0.6 pt, fill=gray, fill opacity=0.2] 
		(1.9,-0.8) -- (2.5,-0.8) -- (2.5,-.3) --(1.9,-0.3) -- cycle; 
    \node[ line width=0.6pt, dashed, draw opacity=0.5] (a) at (0.3,.25)
      {$\sigma_{\mu}$};
         \node[ line width=0.6pt, dashed, draw opacity=0.5] (a) at (2.2,.25)
      {$\sigma_{\nu}$};
          \node[ line width=0.6pt, dashed, draw opacity=0.5] (a) at (0.3,-.55)
      {$K_{a}$};
         \node[ line width=0.6pt, dashed, draw opacity=0.5] (a) at (2.2,-.55)
      {$K_a^{\dagger}$};
		\end{tikzpicture}
	\end{aligned}. \label{eq:EffectTensor}
\end{equation}
Using the effect tensor and correlation tensor, the Born rule becomes $p(a)=\sum_{\mu,\nu}T^{\mu;\nu}E^{\mu;\nu}=\Tr(M_a W)$.

In the two-event case, an intriguing result emerges: the spatial and temporal probability distributions (given by Eqs.~\eqref{eq:pb1} and \eqref{eq:pb2} in  the supplemental material) can both be derived from the same Born rule. Specifically, we can express the joint probability distribution as $p(a,b)=\Tr (M_{a,b}W)$, where $M_{a,b}=K_a\otimes L_b \otimes K_{a}^{\dagger}\otimes L_b^{\dagger}$ and $K_a$ and $L_b$ are the effect operators of the POVMs corresponding to measurement outcomes $a$ and $b$, respectively.
This provides a notable advantage over the pseudo-density operator approach, in which obtaining the temporal measurement statistics requires first solving the Sylvester equation to obtain the quantum channel and initial state, and then substituting them into the spatial and temporal Born rule respectively \cite{fitzsimons2015quantum,fullwood2023quantum}. 
The general spatiotemporal DM and Born rule are illustrated in Fig.~\ref{fig:Born}, where the DET is drawn for joint DM. For local space-time measurement, the DET is the tensor product of the single-event DET in Eq.~\eqref{eq:EffectTensor}

\begin{theorem}
  The spatiotemporal Born rule in the DDO formalism is of the form:
     \begin{align}
      &\text{Local:}\quad   p(a_1,\cdots,a_N)=\Tr ( M_{a_1,\cdots,a_N} W),\\
       &\text{Joint:} \quad p(a)=\Tr(M_a W),
     \end{align}
     where local and joint measurements are defined in definition~\ref{def:measurement}.
Using the doubled correlation tensor $T^{\mu_1,\cdots,\mu_n,\nu_1,\cdots,\nu_n}$ and double effect tensor $E^{\mu_1,\cdots,\mu_n,\nu_1,\cdots,\nu_n}$, the Born rule is given by the contraction of the tensors as shown in Fig.~\ref{fig:Born}.
\end{theorem}

See supplemental material for the proof.
In standard quantum mechanics, joint measurement is exclusively defined for the spatial state, and no similar notion exists for the temporal case. Therefore, the doubled density operator formalism extends beyond standard quantum mechanics. We can also consider arbitrary space-time joint measurements.

\emph{Detecting temporality and causality.} ---
A critical aspect of exploring the space-time state is detecting temporality, which is accomplished in the pseudo-density operator formalism by employing negativity. Regarding the doubled density operator, we obtain the following result:

\begin{theorem}[Criterion of temporality or causality]
    For a given spatially distributed event set $\Acal$, the left or right reduced state of doubled density operator $W_{\Acal}$ must be a density operator. More precisely, $W_{\Acal}^R=\Tr_L W_{\Acal}$ and $W_{\Acal}^L=\Tr_R W_{\Acal}$ are positive semidefinite operators with unit trace.
\end{theorem}

The proof is presented in the supplemental material. This outcome is useful for detecting temporality, as the presence of temporal correlation in $W_{\Acal}$ is indicated when the left or right reduced states are not density operators. Thus, this finding is akin to the positive partial transpose criterion of entanglement \cite{Horodecki2009quantum}.

The utilization of DDOs enables the representation and characterization of space-time quantum processes that manifest indefinite causal order.
See supplemental material for the proof based on causal inequalities.

\emph{Discussion and outlook.} ---
We have demonstrated  a new framework for investigating space-time quantum processes, where space and time are treated on an equal footing. Notably, our framework incorporates a unified form for measurements and the application of the Born rule, thereby establishing a cohesive and integrated approach to the study of these processes.
This paves the way for the generalization of numerous quantum information processes that operate independently in space or time to the unified space-time scenario. 
Typical examples include 
(i) The space-time quantum correlation test, which generalizes the Bell test \cite{bell1964,Brunner2014bell} and Leggett-Garg test \cite{leggett1985quantum,emary2013leggett}. In the space-time scenario, there exists test that is space-time in nature, we have given an example in the supplemental material.
(ii) The communication cost for spatiotemporal quantum correlations which generalizes the communication cost of simulating Bell correlations \cite{Toner2003communicationn} and that of temporal correlations \cite{Brierley2015nonclassicality}. 
(iii) The hierarchy structure of the DDOs that exhibit different spatiotemporal quantum correlations. The hierarchy structure of density operators which exhibit entanglement \cite{Horodecki2009quantum}, steering \cite{Uola2020quantum} and Bell nonlocality \cite{Brunner2014RMP} can be generalized to the temporal DDOs (and more general space-time DDO).
(iv) The DDO offers a valuable avenue for investigating quantum chaos \cite{stockmann2000quantum, hosur2016chaos, Bertini2019exact, foligno2023temporal} and information scrambling within a space-time framework. By utilizing DDOs, we can effectively explore and analyze the complex dynamics and entanglement patterns that arise in space-time scenarios, shedding light on the fundamental aspects of quantum chaos and information scrambling in these contexts.
(v) The DDO also provides a general framework to investigate the quantum  retrodiction and quantum Bayes rule \cite{Leifer2013toward,parzygnat2022axioms}.
(vi) The DDO framework also enables us to study the compatibility of the marginal problems of space-time quantum information processes \cite{jia2023quantumspace}, which include quantum state marginal problem and quantum channel marginal problem as special examples. 
All these topics will be left for our future studies.

\textit{Acknowledgments.} ---
We would like to acknowledge Minjeong Song, Xiangjing Liu, Yixian Qiu, James Fullwood, Arthur Parzygnat and Fabio Costa for valuable communications. 
This work is supported by the National Research Foundation and the Ministry of Education in Singapore.

\let\oldaddcontentsline\addcontentsline
\renewcommand{\addcontentsline}[3]{}
\bibliographystyle{apsrev4-1-title}
\bibliography{mybib}
\let\addcontentsline\oldaddcontentsline

\pagebreak
\clearpage
\widetext
\begin{center}
	\textsf{\large SUPPLEMENTAL MATERIAL\\
 \vspace{1em}
The spatiotemporal doubled density operator: a unified framework for analyzing spatial and temporal quantum processes}
\end{center}
\setcounter{equation}{0}
\setcounter{figure}{0}
\setcounter{table}{0}
\setcounter{page}{1}
\makeatletter
\renewcommand{\theequation}{S\arabic{equation}}
\renewcommand{\thefigure}{S\arabic{figure}}

\tableofcontents

\section{Operational theory for space-time state}

Prior to discussing the space-time doubled density operator 
 (DDO), it is necessary to discuss some key considerations regarding the operational theory for unifying quantum correlations in space and time.
We will utilize a prepare-measure scenario to study quantum correlations.

Let's first introduce the most general framework for a probabilistic physical theory in the prepare-measure scenario, which we call an operational theory.
Consider single-party operational theory, which consists of three ingredients: the state space $\mathsf{State}$; the effect space $\mathsf{Effect}$; and the Born rule
\begin{equation}
    \Pr: \mathsf{State} \times \mathsf{Effect} \to [0,1].
\end{equation}
The operational theory is called \emph{normalised} if there exist two special effects $0,u\in \mathsf{Effect}$ such that
\begin{align}
     \Pr(\rho,0)=0,\quad \forall \rho \in \mathsf{State},\\  \Pr(\rho,u)=1,\quad \forall \rho \in \mathsf{State}.
\end{align}
It's called \emph{state-convex} if $\mathsf{State}$ is a convex subsets of real vector spaces and $\Pr$ is convex-linear on state, i.e., $\forall p\in[0,1], \forall e\in \mathsf{Effect}$, 
\begin{equation}
    \Pr(p\rho+(1-p)\sigma,e)=p \Pr (\rho,e)+ (1-p) \Pr(\sigma,e).
\end{equation}
It's called \emph{effect-complete} if for each $e\in \mathsf{Effect}$ there exists a collection of $e_i\in\mathsf{Effect}$ such that 
\begin{equation}
    \Pr(\rho,e)+\sum_i \Pr(\rho,e_i)=1, \forall   \rho \in \mathsf{State}.
\end{equation}
A set of effects that satisfies the above equality  is called a \emph{measurement}, and the set of all measurements is denoted as $\mathsf{Measurement}$.
Quantum mechanics can be regarded as a specific operational theory, where the states are described by the density operators, effects are described by projectors or more general positive operator-valued measure (POVM), and the Born rule is given by $\Pr(\rho,F_{a|x})=\Tr (F_{a|x}\rho )$.

Now let's consider multiparty operational theory.
Consider $N$ parties $[N]=\{1,\cdots,N\}$, each party has his own operational theory consisting of their states, effects and measurement $\State_i$, $\Effect_i$, $\Measurement_i$ and Born rule $\Pr_i$.
There also is a global operational theory with $\State$, $\Effect$, $\Measurement$ and Born rule $\Pr$, where the joint preparation and joint measurement are allowed.
There is a joint-state map between local states and global state 
\begin{equation}
    J_S: \State_1\times \cdots \times \State_N \to \State
\end{equation}
Similarly, there is a joint-effect map between local effects and global effects
\begin{equation}
    J_E:\Effect_1\times \cdots \times \Effect_N\to \Effect.
\end{equation}
They must satisfy
\begin{equation}\label{eq:local}
\begin{aligned}
    &\Pr (J_S(\rho_1,\cdots,\rho_N),J_E(e_1,\cdots,e_N))\\
    =&{\Pr}_1(\rho_1,e_1)\times \cdots \times {\Pr}_N(\rho_N,e_N),
\end{aligned}
\end{equation}
which means that the local Born rule must be compatible with the global Born rule.
It's also natural to require that the identity effect is preserved: $J_E(u_1,\cdots,u_N)=u$.
In a given space-time, for each event, we have a corresponding operational theory; and for the general event set, there is a global operational theory.
The DDO formalism is an example of such an operational theory.

\section{Spatiotemporal doubled correlation tensor}

In this part, we will show that the spatial, temporal, and spatiotemporal doubled correlation tensors (DCTs) that appear in quantum mechanics satisfy the three conditions in definition \ref{def:STtensor} in the main text.

\emph{Spatial DCT.} --- For the purely spatial case, the proof is straightforward. 
To show that $T^{\{\mu_i\};\{\nu_j\}}$ is Hermitian, notice that 
\begin{equation}
    \begin{aligned}
        {T^{\{\nu_j\};\{\mu_i\}}}^*
        =&(\Tr[ (\otimes_j \sigma_{\nu_j}) \rho (\otimes_i \sigma_{\mu_i})] )^*\\
        =& \Tr [((\otimes_j \sigma_{\nu_j}) \rho (\otimes_i \sigma_{\mu_i}) ])^{\dagger}]\\
        =& \Tr [(\otimes_i \sigma_{\mu_i}) \rho (\otimes_j \sigma_{\nu_j})]=T^{\{\mu_i\};\{\nu_j\}}.
    \end{aligned}
\end{equation}
To show that $T^{\{\mu_i\};\{\nu_j\}}$ is positive semidefinite, consider $X^{\{\alpha_i\}}$, we introduce $G=\sum_{\{\alpha_i\}} (X^{\{\alpha_i\}})^* (\otimes_i \sigma_{\alpha_i})$. Then
\begin{equation}
    \sum_{\{\mu_i\},\{\nu_j\}} {X^{\{\mu_i\}}}^* T^{\{\mu_i\};\{\nu_j\}} X^{\{\nu_j\}}=\Tr (G \rho G^{\dagger})=\Tr ( \rho G^{\dagger}G)\geq 0.
\end{equation}
$T^{0,\cdots0;0,\cdots,0}=1$ is a direct result of the fact $\Tr \rho=1$.

\emph{Temporal DCT.} ---
For the temporal doubled correlation tensor, we need to use the Kraus representations for the quantum channels
\begin{equation}
    \mathcal{E}_l(\bullet)=\sum_{a_l} K^l_{a_l} \bullet ( K^l_{a_l})^{\dagger}.
\end{equation}
A general temporal doubled correlation tensor is of the form
\begin{equation}
    \begin{aligned}
       T^{\{\mu_i\};\{\nu_j\}}= \Tr [\sum_{\{a_l\}}  \sigma_{\mu_n}  K^n_{a_n} \cdots  K^1_{a_1}( \sigma_{\mu_0}\rho \sigma_{\nu_0} )( K^1_{a_1})^{\dagger} \cdots ( K^n_{a_n})^{\dagger}\sigma_{\nu_n}]
    \end{aligned}
\end{equation}
Since each $T_{\{a_l\}}^{\{\mu_i\};\{\nu_j\}}=\Tr[\sigma_{\mu_n}  K^n_{a_n} \cdots  K^1_{a_1}( \sigma_{\mu_0}\rho \sigma_{\nu_0} )( K^1_{a_1})^{\dagger} \cdots ( K^n_{a_n})^{\dagger}\sigma_{\nu_n}]$ is Hermitian, $ T^{\{\mu_i\};\{\nu_j\}}$ is Hermitian.
By introducing $G=\sum_{\{\alpha_i\}} (X^{\{\alpha_i\}} )^*\sigma_{\alpha_n}  K^n_{a_n} \cdots  K^1_{a_1}\sigma_{\alpha_0}$, we see
\begin{equation}
    \sum_{\{\mu_i\},\{\nu_j\}} {X^{\{\mu_i\}}}^* T^{\{\mu_i\};\{\nu_j\}}_{\{a_l\}} X^{\{\nu_j\}}=\Tr (G \rho G^{\dagger})=\Tr ( \rho G^{\dagger}G)\geq 0. 
\end{equation}
This further implies that $T^{\{\mu_i\};\{\nu_j\}}=\sum_{\{a_l\}}T_{\{a_l\}}^{\{\mu_i\};\{\nu_j\}}$ is positive semidefinite.
The $T^{0,\cdots0;0,\cdots,0}=1$ can be derived from the fact that $\Tr \rho=1$ and all $\mathcal{E}_l$ are trace-preserving (TP) maps.

\emph{Spatiotemporal DCT.} ---
The proof for  the general space-time doubled correlation tensor is a combination of the spatial case and temporal case.

\section{Spatiotemporal DDO and generalized Born rule}

For a two-event spatial quantum process, the probability distribution of measuring two projectors $\Pi_a$ and $\Pi_b$ over a bipartite state is of the form
\begin{equation}\label{eq:pb1}
    p_S(a,b)=\Tr ((\Pi_a\otimes \Pi_b) \rho(\Pi_a^{\dagger}\otimes \Pi_b^{\dagger})).
\end{equation}
However, for a two-event temporal quantum process with input state $\rho$ and quantum channel $\mathcal{E}$, the probability distribution of measuring two sequential projectors $\Pi_a$ and $\Pi_b$ is of the form 
\begin{equation}\label{eq:pb2}
    p_T(a,b)=\Tr (\Pi_a \mathcal{E}(\Pi_b\rho\Pi_b^{\dagger}) \Pi_a^{\dagger}).
\end{equation}
Our formalism introduces the corresponding doubled density operators $W_S$ and $W_T$, which are both determined by the doubled correlation tensor for the process. The measurements are represented as $M_{a,b}=\Pi_a\otimes \Pi_b\otimes \Pi_a^{\dagger}\otimes \Pi_b^{\dagger}$, and both spatial and temporal behaviors can be expressed using the same Born rule
\begin{equation}
    p_{S/T}(a,b)=\Tr (M_{a,b} W_{S/T}).
\end{equation}
Furthermore, in addition to unifying the Born rule, our proposed DDO formalism allows for the direct recovery of all physical information. Specifically, (i) for $W_S$, the density operator for the multipartite state can be obtained by taking the reduced state on the left or right half; (ii) for $W_T$, the density operator for each time step can be recovered by taking the reduced state of the corresponding time step; and (iii) when dealing with complicated space-time quantum information process, all the information is encoded in $W_{ST}$, which makes it easier the classify and characterizing different processes. And this also paves the way for studying general space-time quantum correlations rather than dealing with the temporal Legget-Garg scenario \cite{leggett1985quantum,emary2013leggett} and the spatial Bell scenario \cite{bell1964,Horodecki2009quantum,Brunner2014bell} separately.

\emph{Proof of theorem 1.} ---
Consider a general spatial doubled correlation tensor 
\begin{equation}
    T^{\mu_1,\cdots,\mu_N;\nu_1,\cdots,\nu_N}=\Tr[ (\otimes_{i=1}^N \sigma_{\mu_i}) \rho (\otimes_{i=1}^N \sigma_{\nu_i})],
\end{equation}
the corresponding spatial doubled density operator is of the form
\begin{align}
    W=&\frac{1}{d^{2N}}\sum_{\mu_i;\nu_i} T^{\mu_1,\cdots,\mu_N;\nu_1,\cdots,\nu_N} (\otimes_{i=1}^N \sigma_{\mu_i}) \otimes (\otimes_{i=1}^N \sigma_{\nu_i}).
\end{align}
Taking the left partial trace, we obtain
\begin{equation}
    W^R=\frac{1}{d^{N}}\sum_{\nu_i} T^{0,\cdots,0;\nu_1,\cdots,\nu_N}  (\otimes_{i=1}^N \sigma_{\nu_i}).
\end{equation}
Since $T^{0,\cdots,0;\nu_1,\cdots,\nu_N}=\Tr[ (\rho (\otimes_{i=1}^N \sigma_{\nu_i})]$, which is nothing but the Bloch tensor for a multipartite state. This implies  $W^R=\rho$.
Similarly, we can show that $W^L=\rho$.

\emph{Proof of theorem 2.} ---
We need to show that for a general space-time quantum process, the measurement statistics obtained from standard quantum mechanics coincide with Born rule given in theorem 2. 
Here we only consider the local measurements, since for spatial joint measurement, the proof is similar.
For a general space-time doubled correlation tensor 
\begin{align}
T^{\mu^0_{1},\cdots,\mu^0_{m_0},\cdots,\mu^n_{1},\cdots,\mu^n_{m_n};\nu^0_{1},\cdots,\nu^0_{m_0},\cdots,\nu^n_{1},\cdots,\nu^n_{m_n}}=  
\Tr \left[ (\otimes_{j_n} \sigma_{\mu^n_{j_n}})  \mathcal{E}_n (\cdots  \mathcal{E}_1((\otimes_{j_0} \sigma_{\mu^0_{j_0}})\rho  (\otimes_{l_0} \sigma_{\nu^0_{l_0}})) \cdots)  (\otimes_{l_n} \sigma_{\nu^n_{l_n}}) \right] 
\end{align}
the corresponding DDO is of the form
\begin{align}
    W=\frac{1}{d^{2N}}\sum_{\mu_i;\nu_i} T^{\mu_0,\cdots,\mu_N;\nu_0,\cdots,\nu_N} (\otimes_{i=0}^N \sigma_{\mu_i}) \otimes (\otimes_{i=0}^N \sigma_{\nu_i}).
\end{align}
For each local POVM $F_{a_i}^i=(K^i_{a_i} )^{\dagger}K_{a_i}^i$ for $i$-th event, the doubled measurement is of the form
\begin{equation}
    M^i_{a_i}=K^i_{a_i} \otimes (K^i_{a_i} )^{\dagger}
\end{equation}
The measurement statistics in quantum mechanics is given by
\begin{equation}
    p(\{a_i\})=\Tr [
 (\otimes_{j_n} K_{a_{j_n}}^{j_n}) \mathcal{E}_n
  (\cdots  \mathcal{E}_1 ( \otimes_{j_0} K_{a_{j_0}}^{j_0} \rho  ( \otimes_{j_0} (K_{a_{j_0}}^{j_0})^{\dagger} ) ) \cdots )( \otimes_{j_n} (K_{a_{j_n}}^{j_n})^{\dagger} )
    ].
\end{equation}
Since $\{\sigma_{\mu}\}$ form a basis of the operator space $\mathcal{B}(\mathcal{H})$, we can expand each $K_{a_i}^i=\sum_{\mu_i} g_{\mu_i} \sigma_{\mu_i}$ with $g_{\mu_i}=\Tr (K_{a_i}^i \sigma_{\mu_i})/d$.
In this way, we see that $ p(\{a_i\})$ equals the inner product of doubled correlation tensor and doubled effect tensor, which matches well with the expression of Born in theorem 2.

\section{Closed-form expression of doubled density operator}

In this part, we introduce the channel-state duality map in the generalized Pauli basis and discuss how to obtain an explicit expression of doubled density operator using this map.

The channel-state duality is usually characterized by Choi-Jamio{\l}kowski isomorphism in the standard basis of operators\cite{choi1975completely,jamiolkowski1972linear}.
The Choi matrix of a quantum channel $\mathcal{E}$ is of the form
\begin{equation}
    C[\mathcal{E}]=\sum_{i,j}\mathcal{E}(E_{ij})\otimes E_{ij}
    = \mathcal{E}\otimes \operatorname{id} (|\mathds{I}\rrangle\llangle \mathds{I}|).
\end{equation}
The Jamio{\l}kowski matrix is the partial transpose of the Choi matrix
\begin{equation}
    J[\mathcal{E}]=\sum_{i,j}\mathcal{E}(E_{ij})\otimes E_{ji}=  C[\mathcal{E}]^{T_B}.
\end{equation}
Recall that for qubit case, 
\begin{equation}
    U_{\rm swap}=\sum_{i,j=0}^1 E_{ij}\otimes E_{ji}
    =\frac{1}{2}\sum_{\alpha=0}^3 \sigma_{\alpha}\otimes \sigma_{\alpha},
\end{equation}
thus we have 
\begin{equation}
    J[\mathcal{E}]= \mathcal{E}\otimes \operatorname{id} ( U_{\rm swap})=\frac{1}{2} \sum_{\alpha} \mathcal{E}(\sigma_{\alpha}) \otimes \sigma_{\alpha}.
\end{equation}
This inspires the proposal of a novel channel-state duality map in the  generalized Pauli operator basis, which is the Hilbert-Schmidt basis of operators. Notably, in higher dimensions, this map diverges from the Jamio{\l}kowski matrix.

For single qudit state $\rho \in \Herm(\mathcal{H}_I)$, and quantum channel $\mathcal{E}: \mathcal{B}(\mathcal{H}_I) \to \mathcal{B}(\mathcal{H}_O)$, we define the \emph{generalized Jamio{\l}kowski matrix} as
\begin{equation}
    \tilde{J}[\mathcal{E}]=\frac{1}{d}\sum_{\alpha=0}^{d-1}\mathcal{E}(\sigma_{\alpha})\otimes \sigma_{\alpha} \in \mathcal{B}(\mathcal{H}_O) \otimes \mathcal{B}(\mathcal{H}_I).
\end{equation}
Then we have
\begin{equation}
    \mathcal{E}(\rho)=\Tr_{I} \tilde{J}[\mathcal{E}] (\mathds{I}\otimes \rho),
\end{equation}
where the subscript $I$ means the trace is taking over the input space $\mathcal{H}_I$.

For the general $n$-qudit state $\rho$, we define
\begin{equation}
     \tilde{J}[\mathcal{E}]=\frac{1}{d^n}\sum_{\{\alpha_i\}=0}^{d-1}\mathcal{E}(\otimes_i\sigma_{\alpha_i})\otimes(\otimes_i \sigma_{\alpha}) \in \mathcal{B}(\mathcal{H}_O) \otimes \mathcal{B}(\mathcal{H}_I).
\end{equation}
And we similarly have
\begin{equation}
    \mathcal{E}(\rho)=  \Tr_{I} \tilde{J}[\mathcal{E}] (\mathds{I}\otimes \rho).
\end{equation}
It's easy to prove that $\Jtilde[\mathcal{E}]$ is Hermitian since all $\sigma_{\alpha}$'s are Hermitian and $\mathcal{E}$ is Hermicity-preserving.
And the trace of $\Jtilde[\mathcal{E}]$ is $1$ since $\mathcal{E}$ is trace-preserving and $\Tr \sigma_{0}=\Tr \mathds{I}=d$ and $\Tr \sigma_j =0$ for $j\geq 1$.

In the doubled density operator formalism, it's useful to define the \emph{doubled generalized Jamio{\l}kowski matrix}  as (for single particle case) 
\begin{align}
     &\mathcal{J}^{\{\mu_i\};\{\nu_i\}}[\mathcal{E}_1,\cdots,\mathcal{E}_{N-1}]\nonumber\\
    =&\frac{1}{d}\sum_{\alpha=0}^{d-1}[\sigma_{\mu_N} \mathcal{E}_{N-1}(\cdots \mathcal{E}_1(\sigma_{\mu_1}\sigma_{\alpha}\sigma_{\nu_1}) \cdots)\sigma_{\nu_N}] \otimes \sigma_{\alpha}.
\end{align}
   
The doubled correlation tensor can be obtained by 
\begin{equation}
  T^{\{\mu_i\};\{\nu_i\}} = \Tr  \mathcal{J}^{\{\mu_i\};\{\nu_i\}}[\mathcal{E}_1,\cdots,\mathcal{E}_{N-1}] (\mathds{I}\otimes \rho).
\end{equation}
More interestingly, if we define 
\begin{equation}
    \mathcal{W}=\sum_{{\{\mu_i\};\{\nu_i\}}} \mathcal{J}^{\{\mu_i\};\{\nu_i\}}[\mathcal{E}_1,\cdots,\mathcal{E}_{N-1}] \otimes [(\otimes_{i}\sigma_{\mu_i})\otimes (\otimes_j \sigma_{\nu_j})]
\end{equation}
the doubled density operator is given by
\begin{equation}
W=    \Tr_{\mathcal{J}}   \mathcal{W} (\mathds{I}\otimes \rho \otimes \mathds{I} \otimes \cdots \otimes \mathds{I}),
\end{equation}
where the trace for $\mathcal{J}$ means that the trace is taking over the spaces that support $\mathcal{J}^{\{\mu_i\};\{\nu_i\}}[\mathcal{E}_1,\cdots,\mathcal{E}_{N-1}] $.
This gives us a closed-form expression for a general temporal doubled density operator. The generalization to the space-time case is straightforward.

\section{Doubled measurements}

Consider a single-event doubled density operator given by Eq.~\eqref{eq:singleW} in the main text, a measurement is typically modeled by a POVM denoted by $\{F_a\}_a$, where $F_a$'s are a positive semidefinite operator and $\sum_a F_a=\mathds{I}$. A crucial result in quantum measurement theory is that each $F_a$ can be expressed in terms of a pair of operators $K_a$ and its Hermitian conjugate $K_a^{\dagger}$ as $F_a=K_a^{\dagger}K_a$. It should be noted that $K_a$ is not necessarily Hermitian.
The Luder's rule for the post-measurement state is 
\begin{equation}
    \rho\mapsto \frac{K_a \rho K_a^{\dagger}}{\Tr(K_a \rho K_a^{\dagger}) }.
\end{equation} 
The Born rule is of the form
\begin{equation}
    p(a)=\Tr(K_a \rho K_a^{\dagger})=\Tr (F_a \rho).
\end{equation}
Notice that POVM contains projective measurements as special examples, where $F_a=\Pi_a$ and $K_a=\Pi_a=K_a^{\dagger}$.
The doubled measurement (DM) in our DDO formalism is of the form 
\begin{equation}
    M_a= K_a\otimes K_a^{\dagger}.
\end{equation}
The Born rule is given by
\begin{equation}
    p(a)=\Pr (M_a, W)=\Tr (M_aW),
\end{equation}
which is easily checked to match well with the result of standard quantum mechanics. For general multi-event case, the DM is just the tensor product of  single-event DMs, it can also be checked that the measurement statistics obtained from quantum mechanics matches well with that obtained from DDO formalism.

The DDO approach eliminates the need for this additional step and simplifies the overall calculation process.
Furthermore, the usage of the DDO formulation provides a unified and concise expression for the Born rule that is applicable to both spatial and temporal quantum processes. This property renders it particularly well-suited for investigating space-time quantum information processes.

\section{Indefinite causal order}
In order to demonstrate the capability of a DDO to exhibit an indefinite causal order, we examine a basic scenario involving two points in time.
Our analysis employs the one-way signaling framework as a means to characterize the underlying causal order, which is carefully analyzed in Ref. \cite{branciard2015simplest,Abbott2016multi}.

Suppose Alice and Bob are located in a given space-time, we denote $A \prec B$ when Alice's events are before those of Bob's system.
In this case, Bob cannot signal to Alice, the correlation between the measurement outcomes for Alice and Bob must satisfy the one-way signaling condition from Alice to Bob
\begin{equation}
    p^{A\prec B}(a|x,y)= p^{A \prec B}(a|x,y'), \forall a, x, \forall y,y',
\end{equation}
where $x$ is measurement choice of Alice, $y,y'$ are measurement choices of Bob.
Similarly, when $B\prec A$, we have a similar one-way signaling condition from Bob to Alice.
The sets of correlations $p^{A\prec B}$ form a convex set, and similarly for  correlations $p^{B\prec A}$.
A correlation is defined as causally definite if it can be decomposed as 
\begin{equation}
    p(a,b|x,y)=\lambda p^{A\prec B}(a,b|x,y)+(1-\lambda) p^{B\prec A}(a,b|x,y),
\end{equation}
where $0\leq \lambda \leq 1$.
The set of correlations with definite causal orders thus is a convex hull of $A\prec B$ and $B\prec A$ polytope, we denote is as $\mathbb{P}_{\rm DCO}$.
There exist correlations that are outside $\mathbb{P}_{\rm DCO}$, which can be detected via the causal inequalities.
If Alice and Bob both are restricted to choose two dichotomic measurements, these causal inequalities can be determined explicitly \cite{branciard2015simplest,Abbott2016multi}.
They are of the form
\begin{align}
    \frac{1}{4} \sum_{x, y, a, b} \delta_{a, y} \delta_{b, x} p(a, b \mid x, y) \leqslant \frac{1}{2},\\
    \frac{1}{4} \sum_{x, y, a, b} \delta_{x(a \oplus y), 0} \delta_{y(b \oplus x), 0} p(a, b \mid x, y) \leqslant \frac{3}{4},
\end{align}
where $\oplus$ denotes addition modulo 2.

In order to show that there exist DDO and DM that violate the above inequalities,  we will use two results in Ref. \cite{cotler2018superdensity,Silva2014pre}.
In Ref. \cite{Silva2014pre}, a connection between the process matrices and multiple states is established.
It's shown that for any two-party process matrix $\Upsilon$, and measurement setting $M_a,N_b$ (CJ matrices of measurement operator), there exist  bipartite two-time state and corresponding measurements that give the same probability distribution
\begin{equation}
    p(a,b)=\Tr (\Upsilon (M_a\otimes N_b))= \eta_{\Upsilon}\cdot (J_a\otimes K_b),
\end{equation}
where $\eta_{\Upsilon}$ is the two-time state, and $J_a,K_b$ are measurements, `$\cdot$' denote the contraction of the two-time state.
From Ref. \cite[Appendix B.2]{cotler2018superdensity}, a correspondence between the two-time state and superdensity operator is established, where the measurement statistics of the two-time state are also recovered in the superdensity operator formalism.
Combining these two correspondences, we obtain a correspondence between a two-party process matrix and a two-event doubled density operator, where the measurement statistics keep unchanged.
Now map the examples of the process matrices and measurement settings that violated the causal inequalities \cite{branciard2015simplest}, we obtain the DDO and DM that violate the causal inequality.



\section{Testing space-time quantum correlations}

\begin{figure}
    \centering
    \includegraphics[scale=0.7]{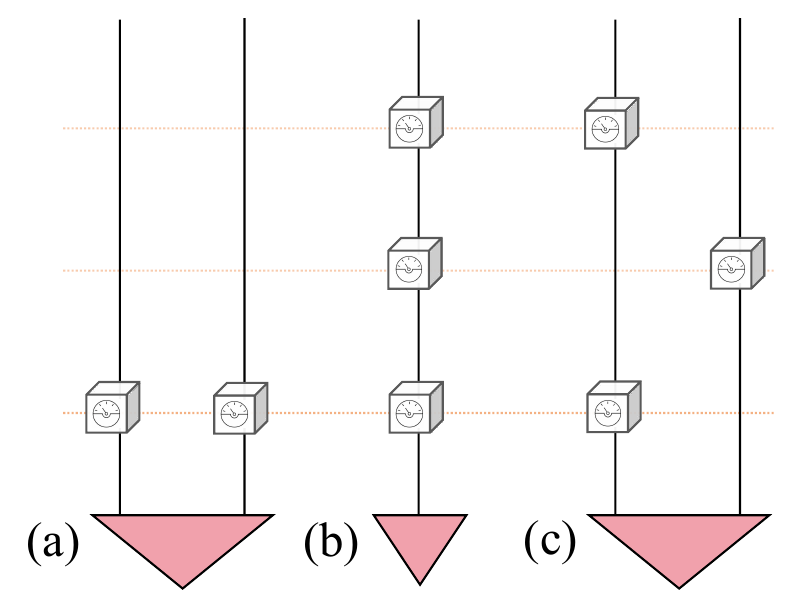}
    \caption{The configurations for space-time quantum correlation test. (a) The spatial quantum correlation test in the Bell scenario, where measurements are implemented on different systems but at the same time slice.
(b) The temporal quantum correlation test in the Leggett-Garg scenario, where measurements are implemented on the same system but at different time slices.
(c) The general space-time quantum correlation test scenario, where measurements are implemented on different systems and at different time slices.}
    \label{fig:ST_test}
\end{figure}

The spatial and temporal quantum correlations are usually investigated in different frameworks, the Bell test for spatial behavior and the Leggett-Garg test for temporal behavior.
The doubled density operator formalism provides a unified framework for investigating quantum correlations exhibited by quantum behavior, which can be either spatial, temporal, or spatial-temporal in nature (See Fig.~\ref{fig:ST_test}). In the spatial case, a state is considered Bell nonlocal if there are measurement settings that violate some Bell inequality \cite{Brunner2014RMP}. However, for the temporal case, traditional approaches have difficulty classifying temporal quantum processes as nonlocal or local, as the violation of the Leggett-Garg inequality may result from either the initial state or the evolution \cite{emary2013leggett}. Using the doubled density operator, this issue can be resolved by defining temporal nonlocal states as those that violate the Leggett-Garg inequality for some measurement settings, since the initial state and evolution information are both encoded in the temporal doubled density operators.
More crucially, in the doubled density operator formalism, we can construct the space-time correlation test that goes beyond the Bell and Leggett-Garg scenario (i.e., the correlation is space-time in nature).


To provide a clearer explanation, let us consider a two-qubit state labeled by $q_1$ and $q_2$. We implement three $\pm 1$-valued measurements at different time slices, $t_1$, $t_2$, and $t_3$. The first measurement, denoted as $Q_1(q_1,t_1)$, is implemented on the first qubit $q_1$ at time $t_1$. The second measurement, denoted as $Q_2(q_2,t_2)$, is implemented on the second qubit $q_2$ at time $t_2$. Finally, the third measurement, denoted as $Q_3(q_1,t_3)$, is implemented on the first qubit $q_1$ at time $t_3$.
Using the assumptions of space-time realism and space-time locality, we can make the following statements:
\begin{equation}\label{eq:ST_test}
    \langle Q_2 Q_1\rangle +\langle Q_3Q_2 \rangle -\langle Q_3 Q_1 \rangle \leq 1.
\end{equation}
In contrast, quantum mechanics predicts a violation of this inequality. Let us consider a scenario where two systems share a singlet state $|\psi^-\rangle$, and three measurements with Bloch vectors $\vec{a}_i$, $i=1,2,3$ are implemented at different time slices. Since the measurements $Q_1$ and $Q_2$ are implemented on different qubits of the singlet state, we have $\langle Q_2 Q_1\rangle =-\vec{a}_2\cdot \vec{a}_1$. Similarly, the measurements $Q_3$ and $Q_2$ are also implemented on different qubits, yielding $\langle Q_3Q_2 \rangle=-\vec{a}_3\cdot \vec{a}_2$. On the other hand, the measurements $Q_1$ and $Q_3$ are implemented on the same qubit, which leads to a reduced state for $q_1$ that is the maximally mixed state $\rho=\mathds{I}/2$, and thus $\langle Q_3 Q_1 \rangle=\vec{a}_3\cdot \vec{a}_1$. Denoting the angle between $\vec{a_i}$ and $\vec{a_j}$ as $\theta_{ij}$, we see that quantum mechanics can reach a value of $-\cos \theta_{12}-\cos \theta_{23}+\cos \theta_{13}$ for this inequality. By choosing $\theta_{13}=0$ and $\theta_{12}=\theta_{23}=\pi$, we find that quantum mechanics can achieve a maximal violation of $3$ for the inequality. It is worth noting that although Eq.~\eqref{eq:ST_test} has a similar form to the Leggett-Garg inequality, they are fundamentally different. The maximum quantum violation for the space-time test exceeds that of $3/2$ for the Leggett-Garg inequality. This difference can be seen clearly from the expression of doubled density operators, where one is a purely temporal doubled density operator and the other is a space-time doubled density operator.


\section{Dynamics of DDO}

We have defined the spatiotemporal state as the DDO, and a natural question arises: how do we define the dynamics of the DDO? Since the quantum causal structure is encoded in the DDO, its dynamics also characterize the dynamics of the quantum causal structure. The motivation for understanding the dynamics of quantum causal structure stems from quantum gravity. In general relativity, the causal structure is considered to be dynamical, and when we attempt to quantize it, the corresponding quantum causal structure must also exhibit dynamical behavior.
Here we develop a framework for the dynamics of quantum causality by introducing transformations between DDOs, which we refer to as the doubled quantum channel (DQC).
The DQC allows us to study how quantum causal structures evolve and how information flows within a spatiotemporal system. By analyzing the transformations between DDOs, we gain insights into the dynamics of quantum causality and the underlying mechanisms that govern the evolution of quantum systems.

\begin{definition}[Doubled quantum channel]
For two event sets $\Acal$ and $\Bcal$, we define the doubled quantum channel (DQC) as a complex linear map
\begin{equation}
    \Phi: \mathcal{B}(\mathcal{H}_{\Acal}^L\otimes \mathcal{H}_{\Acal}^R) \to \mathcal{B}(\mathcal{H}_{\Bcal}^L\otimes \mathcal{H}_{\Bcal}^R)
\end{equation}
such that DDOs are mapped to DDOs.
Since a DDO is equivalently characterized by a DCT $T^{\{\mu_i\};\{\nu_j\}}$, a DQC can equivalently be characterized by a tensor $\Phi^{\{\alpha_k\};\{\beta_l\}}_{\{\mu_i\};\{\nu_j\}}$ such that  for any $T^{\{\mu_i\};\{\nu_j\}} \in \mathsf{DCT}(N)$, the resultant
\begin{equation}
    R^{\{\alpha_k\};\{\beta_l\}}=\sum_{{\{\mu_i\};\{\nu_j\}}} \Phi^{\{\alpha_k\};\{\beta_l\}}_{\{\mu_i\};\{\nu_j\}} T^{\{\mu_i\};\{\nu_j\}}
\end{equation}
satisfies $R^{\{\alpha_k\};\{\beta_l\}} \in \mathsf{DCT}(M)$.
\end{definition}

It is important to note that the DQC, when represented as a matrix, must satisfy certain properties. Firstly, it should be Hermicity-preserving with respect to both left and right indices. Additionally, it must be a positive map with respect to these indices. Moreover, the resulting DCT, after applying the DQC, should have an element $R^{{0,\cdots, 0};{0,\cdots,0}}$ equal to 1 for all input DCTs. Although it is natural to impose the condition that all single-event reduced DCTs of the resultant DCT follow the form described in Eq.~\eqref{eq:singleT} in the main text, we did not explicitly adopt this condition as an axiom of the DCT in this work, so it can be omitted here.

\begin{example}
In the special case where both event sets $\Acal$ and $\Bcal$ correspond to fixed quantum processes, namely, the quantum process of $\Acal$ is represented by $(\rho, \mathcal{E}_1,\cdots,\mathcal{E}_n)$, and the quantum process of $\Bcal$ is represented by $(\omega,\mathcal{F}_1,\cdots,\mathcal{F}_n)$, the DQC can be realized through a quantum channel $\mathcal{L}$ that maps $\rho$ to $\omega$, along with $n$ higher-order maps $\Psi_k$ that map channels to channels, where $\Psi_k:\mathcal{E}_k\mapsto \mathcal{F}_k$ for $k=1,\cdots,n$.
\end{example}

\end{document}